# Spatial-resolved X-ray photoelectron spectroscopy of Weyl semimetal NbAs

H W Liu[1], G H Zhang[2], P Richard[1,3,4], L X Zhao[1], G F Chen[1,3,4] and H Ding[1,3,4]

[1] Institute of Physics, Chinese Academy of Sciences, Beijing 100190, China

[2] Dalian Institute of Chemical Physics, Chinese Academy of Sciences, Dalian 116023, Liaoning, China

[3] School of Physics, University of Chinese Academy of Sciences, Beijing 100190, China

[4] Collaborative Innovation Center of Quantum Matter, Beijing, China

E-mail: hliu@iphy.ac.cn and dingh@iphy.ac.cn

**Abstract**

We utilized X-ray photoemission electron microscopy (XPEEM) and X-ray photoelectron spectroscopy (XPS) to investigate the crystal surface of Weyl semimetal NbAs. XPEEM images present white and black contrast in both the Nb 3*d* and As 3*d* core level spectra. Surface-sensitive XPS spectra indicate that the entire surface of the sample contains both surface states of Nb 3*d* and As 3*d*, in form of oxides, and bulk states of NbAs. Estimated atomic percentage values $n_{Nb}/n_{As}$ suggest that the surface is Nb-rich and asymmetric for white and black areas.

**Introduction**

In 1929, H. Weyl proposed that the massless solution of the Dirac equation represents a pair of new particles, the so-called Weyl fermions [1]. Their existence in particle physics remains elusive even after more than eight decades. However, theorists recently predicted that the Weyl fermions exist in the TaAs family of crystalline compounds (TaAs, NbAs, NbP, and TaP) [2], which was followed by their experimental observation by angle-resolved photoemission spectroscopy (ARPES) [3-7]. Due to inversion symmetry-breaking, the surface electronic structure of these materials is characterized by an arclike topology connecting the surface projection of 24 Weyl nodes in the bulk band structure [2]. The discovery of the Weyl semimetals has received worldwide interest and it is believed to open a new area in condensed matter physics after graphene and the three-dimensional topological insulators.

The surface structure of the TaAs family of compounds is not fully understood. In principle, both (Ta,Nb) and (As,P) surface terminations are as likely to occur. This is consistent with Souma *et al.*

claiming that the cleaved surface of NbP is a single domain with either a Nb or P surface termination [8]. In TaP, both types of surface terminations are also detected but they coexist, indicating that the cleaved surface is not a single domain [9]. In contrast, Lv *et al.* report that for a cleaved surface of TaAs, only the As-termination is observed by ARPES [3].

Motivated by the lack of understanding of the surface structure of the TaAs family of compounds, which is important to their topological properties, here we report a study of spatial-resolved X-ray photoemission electron microscopy (XPEEM) and X-ray photoelectron spectroscopy (XPS) of Weyl semimetal NbAs.

**Experimental**

The NbAs plate-like samples we studied were grown by chemical vapour transport. The crystal orientation has been determined by X-ray diffraction [10]. The NbAs crystal has a non-centrosymmetric structure corresponding to space group $I4_1md$ (109) and point group $C_{4v}$, as shown in Fig. 2(a). We utilize a commercial spectroscopic photoemission and low-energy electron microscope (SPELEEM from Elmitec Co. Ltd) at the XPEEM endstation of the Dreamline (beamline 09U) at Shanghai Synchrotron Radiation Facility to record mirror electron microscopy (MEM), XPEEM images and spatial-resolved XPS. The angle between the X-ray beam and the sample surface is 16 degrees. Samples with a typical size of $0.4 \times 0.4 \times 0.08$ mm$^3$ were cleaned in an ultrasonic bath, transferred in air into the SPELEEM chamber, and then annealed at 300℃ for 8 to 10 hours to remove the adsorbates. We notice that all attempts to measure cleaved samples failed due to sample arcing. During the XPEEM measurements the typical pressure was below $3 \times 10^{-9}$ torr.

**Results and discussion**

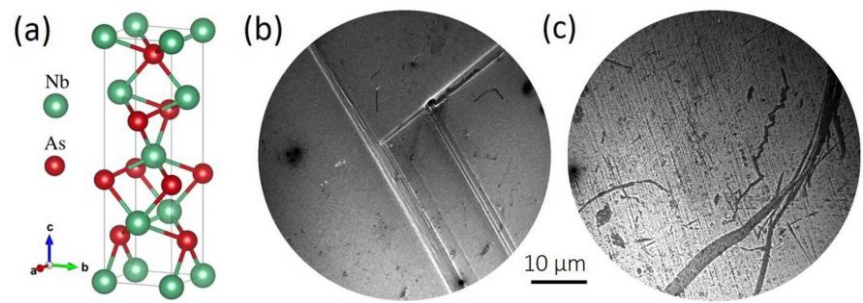

**Figure 1.** (a) Molecular structure of NbAs. (b) and (c) Typical MEM images.

MEM images reveal the surface potential as determined by topography, work function and surface charge variations [11]. Taking advantage of higher lateral resolution, Fig. 1b shows typical natural growth tracks with rectangular shape bundle steps on the NbAs surface, which are consistent with the (001) facet of NbAs. Dense and parallel segments and sharp contrast of patterns can be seen in Fig.

1(c), indicating that the surface is spatial-resolved. We note that the surface does not produce low-energy electron diffraction pattern, suggesting that it is amorphous.

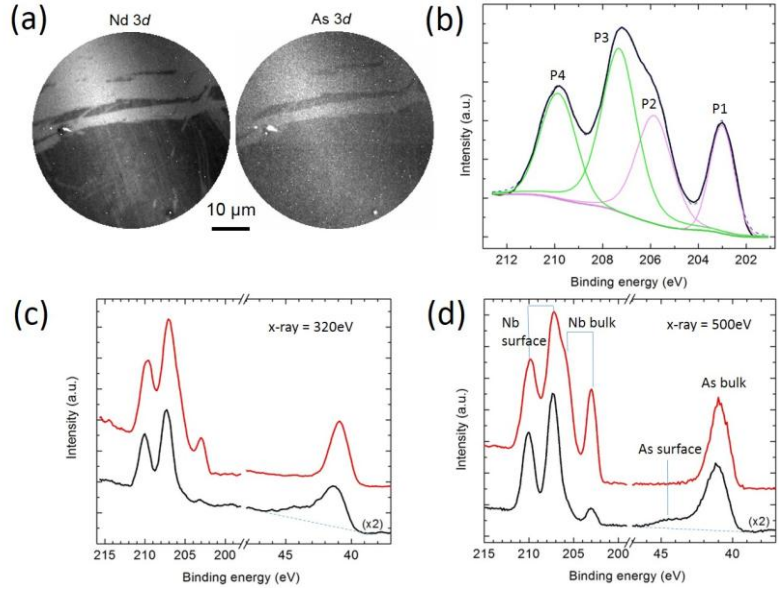

**Figure 2.** (a) Typical XPEEM maps of NbAs sample. (b) Typical XPS spectrum of Nb 3*d* core level. Two sets of peaks, P1/P2 and P3/P4, are resolved. (c) Typical 3*d* core levels spectra of Nb and As recorded with 320 eV photons. (d) Same as (c) but using 500 eV photons. The red curves in (c) and (d) have been extracted from the white area whereas the black curves come from the black area.

The PEEM technique detects electrons emitted from atoms with kinetic energy $E_{kin} = h\nu - E_{bin} - \varphi$, where $h\nu$ is the photon energy and $\varphi$ the work function [12]. Typically, $h\nu$ is fixed and $E_{kin}$ is varied, which allows probing the chemical states of the emitting atoms by measuring the binding energies of their core electrons. The intensity of the photoemission signal is proportional to the number of emitters in the topmost layers within their energy-dependent escape depth, and thus provides straightforward and quantitative information about the chemical composition at the surface.

Figure 2a compares average XPEEM maps for kinetic energies tuned around the Nb 3*d* and As 3*d* core levels. The dark/bright contrast is inversed in MEM images (not shown here). The emitted flux and brightness contrast change with energy in the XPEEM maps. The maps recorded with energies close to the core levels have sharper brightness contrast. In our measurements, the image contrast for Nb is always sharper than for As, which may be caused by different photoemission cross sections or sensitivity factors. White and black areas coexist on the surface and the patterns do not follow the underlying topographies.

A typical XPS spectrum near the Nb 3*d* binding energy is shown in Fig. 2b. The binding energy of Nb $3d_{5/2}$ electrons in NbAs is found between 202 and 204 eV, as compared to 202.0 eV in metallic Nb and 203.8 eV in NbN. Four peaks can be resolved, as illustrated by a Lorenztian fit. This corresponds to twice the number of peaks expected for bulk NbAs, indicating the existence of a second Nb site. Similarly, previous ARPES studies reported two As sites, as deduced by the observation of 4 core level peaks [3,5,6]. Here we detect the As 3$d$ core levels at 40.9 eV (Figs. 2c and 2d). Due to a lower energy resolution than in the ARPES measurements, we can resolve clearly neither the 3$d_{3/2}$ and 3$d_{5/2}$ components, nor additional splitting. However, the peak shape is irregular, indicating that more than one peak is present.

In order to understand the nature of the two Nb sites, we use the PEEM capability of resolving the evolution of the Nb 3$d$ core levels with respect to space. The P3 and P4 peaks at higher binding energy are observed systematically at every points and for all photon energies used. The intensity of the accompanying P1 and P2 (a shoulder) peaks at lower energy, however, strongly depends on the x-ray energy and local areas, and it sometimes nearly vanishes. More specifically, the P1 and P2 peaks have much weaker intensity in the dark areas in the XPEEM images in Fig. 2(a). This is well illustrated by the spectra displayed in Figs. 2(c) and 2(d). We thus assign the four peaks to two pairs: 203 eV (P1) / 205.8 eV (P2), and 207.2 eV (P3) / 209.9 eV (P4). Each pair has an energy splitting of ~2.7 eV, which is consistent to our expectation for the spin-orbit-split 3$d_{5/2}$ and 3$d_{3/2}$ levels of Nb 3$d$ core levels.

Our results suggest that the P1 and P2 peaks are related to bulk NbAs states while the P3 and P4 peaks, less bulk-sensitive, are associated to photoemission of Nb at the surface. To support this assumption, we performed XPS measurements from the same selected areas after adjusting the incident x-rays to 660, 580, and 480 eV. In the soft X-rays regime, the bulk-sensitivity of the photoemission spectroscopies increases with photon energy. As shown in Fig. 3(a), the spectral weight of the P1 peak evolves oppositely to that of peak P4, with the intensity of the P1 and P4 peaks increasing and decreasing with photon energy, respectively. Not only this confirms that the two peaks correspond to different Nb sites, it strongly suggests that the P1/P2 pair is associated with the bulk whereas the other pair is related to the surface. This observation is valid for both the white and black areas.

A contrast also exists between the As spectra recorded in the white areas, for which the bulk peaks of Nb are enhanced, and the black areas. Notably, the intensity of the As peak is reduced in the black areas. At the same time, the full-width-at-half-maximum of this peak increases from 1.8 to 2.2 eV, with the peak enlargement occurring on the high-energy side, where the bulk components of the As core levels are reported in ARPES experiments [5]. In other words, and contrary to what we observe for the Nb peaks, the As surface states rather than the bulk states are mainly affected by the switch from the white to the black area.

Although we detect clearly the bulk states of NbAs, the interpretation of the P3 and P4 peaks in terms of the topological NbAs surface states is rather difficult to reconcile with our data. Here we recall that the surface studied, though cleaned, is natural and it is thus subject to contamination, notably by oxygen. Nb can form different oxides, each having their own signature of Nb core levels [13]. The energies of the P3 and P4 peaks correspond well to the Nb 3$d_{5/2}$ and 3$d_{3/2}$ core levels in $Nb_2O_5$ [14]. Interestingly, we observe a small bump at a binding energy of 44.6 eV that is also consistent to the 3$d$ core levels of As in $As_2O_3$ [14]. We believe that oxide layers of either $Nb_2O_5$ or $As_2O_3$ form in the short period before the sample insertion into the vacuum chamber. We also notice from Fig.2c and 2d that the weak bulk state of Nb 3$d$ and the surface state of As 3$d$ in $As_2O_3$ are preferable in dark areas, indicating that the black areas are more oxidized than the white area. We conclude that for both Nb and As in NbAs, there are a bulk state at lower binding energy and a surface state at higher binding energy in the form of the oxides, which we suppose due to the binding of the free dangling bonds on the topmost surface with oxygen from air. The existence of an oxide is unexpected in our experiment, in contrast to with ARPES data obtained from cleaved surfaces. It seems that annealing at 300 ℃ is not capable to fully evaporate the oxidized layers. By now, literature lacks about knowledge on Weyl semimetals surface and their annealing.

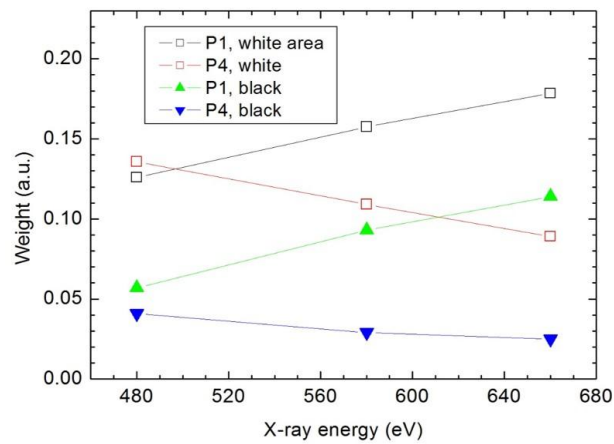

**Figure 3.** Spectral weight of the Nb core-level peaks P1 and P4 as a function of photon energy.

In Fig. 2c, the kinetic energy of photoelectrons of Nb is about 110 eV. The corresponding mean free path of photoelectrons is about 5 Å. There is almost no bulk Nb signal present in the black area, suggesting that the black area is covered by an oxide which is thicker than 5 Å. In the white area, a non-oxidized Nb peak appears and the spectral weight of the Nb oxide becomes smaller than that in the black area. It indicates that the Nb oxide is thinner in the white area. Moreover, the kinetic energy ~280 eV of As photoelectrons in Fig. 2c is comparable to the kinetic energy ~290 eV of Nb photoelectrons in Fig. 2d. However, the spectral weight of the As oxide in the As 3$d$ core level spectrum in Fig. 2c is much weaker than that of the Nb oxide in the Nb 3$d$ core level spectrum in Fig. 2d, indicating that As in the black area is less oxidized, as same as in the white area. Thus As on the surface mainly exists as NbAs bulk. Arsenide bulk intensity is much stronger in the white area than in the black area in Fig.2c and 2d, suggesting that the amount of NbAs is larger in the white area than in the black area.

To generate atomic percentage values ($n_{Nb}/n_{As}$) on the surface, each raw XPS signal must be corrected by dividing the signal intensity ($I_{Nb}$ and $I_{As}$) by a relative sensitivity factor ($S_{Nb} = 2.517$ and $S_{As} = 0.570$) [15]:

$n_{Nb}/n_{As} = (I_{Nb}/S_{Nb})/(I_{As}/S_{As})$.

Our analysis suggests that $n_{Nb}/n_{As}$ is 53.2% : 46.8% for the white area, and 57.8% : 42.2% for the black area. The result is unexpected because the entire area is Nb-rich as well as non-equivalent for the white and black areas. Possibly the calculation is distorted by introduction of oxygen into the topmost surface and the underlying few layers in the form of $Nb_2O_5$ and $As_2O_3$, or due to the difference of the oxidation strength, or somehow missing As atoms from there, for example during evaporation. Though the surface may be oxidized, all previous experiments on the same batches of samples, using different techniques, suggest the good quality of the NbAs bulk [10]. The asymmetry in the elemental composition implies the possibility of a co-existence of As-terminated and Nb-terminated surfaces.

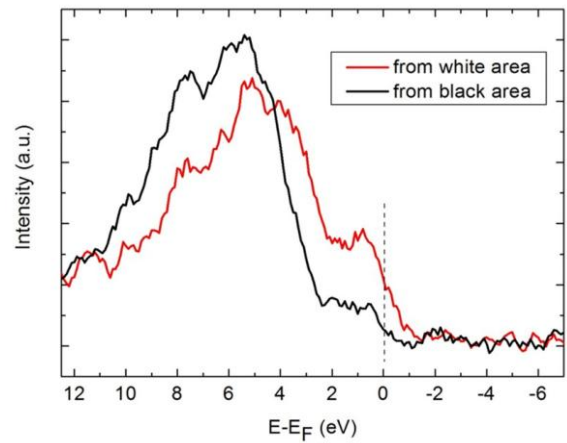

**Figure 4.** Comparison of the near-$E_F$ spectra recorded with 320 eV photon energy in the white (red curve) and black (black curve) areas.

Now that we have identified two types of regions at the sample surface (black and white areas), it is interesting to see what the corresponding low-energy states are, and thus we recorded complementary spectra near the Fermi level. In Figure 4, we compare the spectra associated to both regions up to 12 eV of binding energy. Although the spectra look roughly similar, a clear spectral weight transfer from the low to the higher binding energies is observed for the black area as compared to the white one. In particular, the spectral intensity at the Fermi level is clearly weaker for the black region. This observation supports the formation of insulating oxide at the surface in the black region, for which the spectral weight would be naturally moved away from the Fermi level.

**Summary**

XPEEM and XPS have been utilized to investigate the spatial-resolved electronic structure of Weyl semimetal NbAs surface without cleavage. XPEEM images present white and black contrasts for both the Nb 3*d* and As 3*d* images. Surface-sensitive XPS spectra show that the entire sample surface contains both surface states in form of oxides and NbAs bulk states for both Nb 3*d* and As 3*d*. Estimated atomic percentage values $n_{Nb}/n_{As}$ suggest that the surface is Nb-rich.

**Acknowledgments**

H. W. Liu is grateful to B. Q. Lv and N. Xu for fruitful discussions. This work was supported by grants from NSFC (11674371 and 21403222) of China.